%% file: main.tex
\documentclass{article}

\usepackage[T1]{fontenc}
\usepackage{iclr2027_conference,times}
\usepackage{amsmath,amssymb}
\usepackage{graphicx,wrapfig}
\usepackage{booktabs,tabularx,array,multirow}
\usepackage[table]{xcolor}
\usepackage[most]{tcolorbox}
\usepackage{listings}
\usepackage{capt-of}
\usepackage{url}
\usepackage[hidelinks]{hyperref}

\iclrfinalcopy

\definecolor{HeatBlue}{HTML}{93B5CF}
\definecolor{HeatPale}{HTML}{F1F6FA}
\definecolor{TableAccent}{HTML}{254B70}
\definecolor{OverallRow}{HTML}{EEF1F4}

\newcommand{\heat}[1]{%
  \cellcolor{HeatBlue!#1!HeatPale}\textcolor{black}{#1}%
}

\newcommand{\benchmark}{\textsc{MMSkillRisk}}
\newcommand{\skillfile}{\texttt{SKILL.md}}

\hypersetup{
  pdftitle={MMSkillRisk: Can Agents Stay Safe When Multimodal Skills Become Traps?},
  pdfauthor={Lingqi Jiang, Jialuo Chen, Jianan Ma, Xinhao Deng, Xiaohu Du, Sibo Yi, Yuqi Qing, Zhenguang Liu, Qinming He, Shiwen Cui, Changhua Meng}
}

\title{\centering
MMSkillRisk: Can Agents Stay Safe When\\
Multimodal Skills Become Traps?}

\author{%
\normalfont
\begin{tabular}[t]{@{}c@{}}
Lingqi Jiang$^{1,2}$ \quad
Jialuo Chen$^{1,2,*}$ \quad
Jianan Ma$^{2,3}$ \quad
Xinhao Deng$^{2,4}$ \quad
Xiaohu Du$^{2}$ \\[4pt]
Sibo Yi$^{4}$ \quad
Yuqi Qing$^{4}$ \quad
Zhenguang Liu$^{1,*}$ \quad
Qinming He$^{1}$ \quad
Shiwen Cui$^{2}$ \quad
Changhua Meng$^{2}$ \\[8pt]
$^{1}$Zhejiang University \quad
$^{2}$Ant Group \\[3pt]
$^{3}$Hangzhou Dianzi University \quad
$^{4}$Tsinghua University \\[6pt]
{\small $^{*}$Corresponding authors.}
\end{tabular}%
}

\begin{document}

\maketitle

\fancyhead{}
\renewcommand{\headrulewidth}{0pt}

\input{sections/abstract}
\input{sections/introduction}
\input{sections/related_work}
\input{sections/background}
\input{sections/construction}

\input{sections/experiment}
\input{sections/conclusion}



\subsection*{Ethics Statement}

We develop \benchmark{} to understand and improve
the security of multimodal agents. All experiments use isolated
task environments with synthetic inputs and instrumented tools.
Data-transfer effects are evaluated through a local receiver,
while file modifications and persistence effects are confined
to the experimental environments. We recognize that the attack
techniques could be misused. Our aim is to support security
evaluation and the development of defenses against malicious
multimodal skills.


\bibliography{iclr2027_conference}
\bibliographystyle{iclr2027_conference}

\clearpage
\appendix
\input{sections/appendix}

\end{document}

%% file: sections/abstract.tex
\begin{abstract}
Agent skills are shareable packages of procedural instructions, tools, and examples. Multimodal skills additionally include visual references that agents retrieve and inspect during execution. Because these images guide actions, attackers can disguise malicious instructions as ordinary visual guidance within otherwise legitimate skills. Existing skill-security research primarily examines text-carried attacks or scanner detection, leaving the runtime effects of image-borne attacks insufficiently evaluated.
We introduce \benchmark{}, to our knowledge the first publicly available benchmark dedicated to end-to-end safety evaluation of image-borne attacks in multimodal skills. To instantiate this attack surface, we design Native-Context Visual Attack (NCVA), which disguises malicious instructions as native components of teaching images, such as annotations and interface labels. The accompanying \skillfile{} provides auxiliary guidance toward relevant visual regions without explicitly stating the malicious operation.
Built from 28 curated clean skills, \benchmark{} contains 36 attack packages and 108 executable cases spanning five attack objectives, with separate checks for attack success and legitimate-task completion.
Across nine model--harness configurations evaluated in isolated sandboxes, NCVA induces unauthorized operations in every configuration. Its pooled attack success rate (ASR) reaches 43.1\%, exceeding the matched text-carrier baseline by 16.4 percentage points, with higher ASR in all nine configurations. Attack success and legitimate-task completion co-occur in 36.5\% of cases, reaching 72.2\% for GPT-5.6-sol with Codex.
These results show that skill-bundled images can induce unauthorized actions even as agents complete legitimate tasks, so task success alone does not establish safe skill use. Our code and data are available at \url{https://github.com/kaill-jlq/MMSkillRisk}.
\end{abstract}

%% file: sections/introduction.tex
\section{Introduction}
\label{sec:introduction}

Language-model agents increasingly rely on reusable skills: portable packages of procedural instructions, tools, scripts, and examples that can be shared through open marketplaces and installed to provide specialized capabilities at inference time without retraining~\citep{li2026skillsbench,liu2026agent,xu2026agent}. This convenience also introduces a supply-chain risk: third-party skills installed without careful review can give attackers a foothold to influence agent planning and execution~\citep{schmotz2026skill,jin2026skillsafetybench,liu2026openskillrisk}.

Textual procedures alone, however, cannot always faithfully capture task-relevant information such as chart details, interface geometry, spatial relations, and observable state changes~\citep{xu22026agent,jiang2026visualskill}.
A growing line of multimodal skill systems therefore supplements textual instructions with screenshots, keyframes, and other visual references~\citep{zhang2026mmskills,jiang2026visualskill,xu22026agent,fan2026resource2skill}.
These images are not merely illustrative: agents retrieve and inspect them during execution to determine where to act, when a step applies, and how to interpret an operation's outcome.
Their contents can therefore become a basis for action, creating an opportunity for attackers to disguise malicious instructions as ordinary visual guidance within an otherwise legitimate skill.

Despite this execution-relevant role of visual resources, existing skill-security benchmarks primarily evaluate attacks carried by textual instructions, scripts, and manifests~\citep{schmotz2026skill,jin2026skillsafetybench,liu2026openskillrisk,zhuang2026agenttrap,jiang2026harmfulskillbench}.
The risks introduced by images bundled with skills consequently remain underexplored.
Although \citet{jia2026seeing} examines instructions hidden in skill images, its focus is on detection by security scanners rather than their effects on agents executing legitimate tasks.
Scanner evasion alone does not establish whether an agent will follow the malicious content and carry out an unauthorized operation.
Conversely, successful task completion does not establish safety, because an agent may fulfill the user's request while also performing attacker-directed actions.
What is missing is an end-to-end benchmark that makes image-borne skill attacks executable and measures their security consequences separately from legitimate-task outcomes.
 
To address this gap, we introduce \benchmark{}, to our knowledge the first publicly available benchmark dedicated to end-to-end safety evaluation of image-borne attacks in multimodal skills.
To systematically instantiate this attack surface, we design \textbf{Native-Context Visual Attack} (NCVA).
Rather than appending a visibly separate malicious instruction to an image, NCVA disguises it as a native component of the teaching material, such as an annotation, or an interface label.
The malicious content thus appears to belong to the legitimate procedure the agent follows.
The accompanying \skillfile{} provides auxiliary guidance toward the relevant visual regions without explicitly stating the malicious operation.
This design tests whether agents treat visual reference content as authoritative procedural guidance and translate the embedded instructions into actual actions.

We construct \benchmark{} through a systematic process of clean-skill curation, attack-package construction, and executable case instantiation.
We begin with 28 carefully curated clean skills selected from public skill packages, resources released with prior multimodal skill work~\citep{zhang2026mmskills,che2026mmg2skill}, and visual workflows we authored.
Building on these skills, we construct 36 NCVA attack packages spanning five attack objectives: data exfiltration, persistence, privilege escalation, artifact pollution, and integrity destruction.
We then instantiate 108 evaluation cases, each pairing a legitimate user task with an attacker objective and providing the required inputs, tools, and separate checks for attack success and task completion.
This construction links curated skill workflows to controlled visual attacks and executable outcome checks, enabling systematic evaluation of both unauthorized behavior and its coexistence with successful task execution.

We use \benchmark{} to evaluate nine model--harness configurations, with each case executed in an isolated sandbox. The results show that (i) NCVA yields a pooled ASR of 43.1\%, exceeding the matched text-carrier baseline by 16.4 percentage points, with higher ASR in all nine; (ii) 36.5\% of evaluation cases exhibit both attack success and legitimate-task completion, reaching 72.2\% for GPT-5.6-sol with Codex.

Our contributions are:
\begin{itemize}
    \item We introduce \benchmark{}, a systematically constructed, runnable safety benchmark targeting image-borne attacks in multimodal skills. Built from 28 carefully curated clean skills, it contains 36 attack packages and 108 executable cases, with separate checks for unauthorized behavior and legitimate-task completion based on execution traces and delivered artifacts.
    \item We design NCVA to instantiate this attack surface by embedding malicious instructions as native-looking components of teaching images, while keeping explicit malicious instructions out of the accompanying skill text. This provides a controlled way to evaluate whether agents convert visual reference content into unauthorized actions.
    \item We conduct an end-to-end evaluation across nine model--harness configurations, observing successful visual attacks and higher ASR than the matched text-carrier baseline in all nine. Attack success also frequently coexists with legitimate-task completion, highlighting a security risk: skill-bundled images can induce agents to perform unauthorized actions even as they complete legitimate tasks.
\end{itemize}

%% file: sections/related_work.tex
\section{Related Work}
\subsection{Multimodal Agent Skills}

Agent skills support procedural reuse across tasks. Voyager, for example, maintains a growing library of executable programs that can be retrieved and composed for new tasks~\citep{wang2023voyager}. For visual agents, reusable knowledge also concerns the observations that determine when and how a procedure applies. Recent work therefore explores how to represent and access visual guidance within skills. MMSkills couples textual procedures with state cards and multi-view keyframes, consulting selected visual evidence in a temporary branch to align it with the current environment~\citep{zhang2026mmskills}. VISUALSKILL organizes application-specific knowledge into a hierarchy of topics and provides a tool for loading the relevant text and figures on demand~\citep{jiang2026visualskill}. AutoVisualSkill distinguishes static visual references, dynamic visual working memory, and interleaved instructions that associate individual steps with supporting visual evidence~\citep{xu22026agent}. These approaches give visual resources explicit roles in interpreting states, applying procedures, and tracking progress.

A complementary line of work studies how reusable skills are acquired and updated. Mirage-1 abstracts interaction trajectories into execution, core, and meta-skills, using this hierarchy to guide planning and online exploration~\citep{xie2025mirage}. XSkill extracts both action-level experiences and task-level skills from agent rollouts, grounding their accumulation and retrieval in visual observations~\citep{jiang2026xskill}. Beyond interaction traces, MMG2Skill converts existing multimodal guides into editable skills and refines them using feedback from agent-visible trajectories~\citep{che2026mmg2skill}. RESOURCE2SKILL draws on tutorial videos, repositories, articles, and reference artifacts to construct a hierarchical skill wiki containing text, code, and visual examples, which agents retrieve and compose during execution~\citep{fan2026resource2skill}. 

\subsection{Agent Skill Security}

Public skill repositories contain vulnerable and malicious packages, making third-party skills a supply-chain attack surface~\citep{liu2026agent,liu2026malicious}. Prior attacks place malicious instructions in helper scripts, documentation examples, or the skill body~\citep{jia2026skillject,qu2026supply}. Benchmarks study injection through skill-related interfaces and test whether an agent carries out the attack~\citep{schmotz2026skill,jin2026skillsafetybench,guo2026malskillbench}. SkillCamo is closest to our setting: it hides instructions in skill images and rewrites the surrounding documentation to bring those images into the workflow~\citep{jia2026seeing}.

Defenses inspect skill files through static or semantic analysis~\citep{cisco2026skillscanner,bhardwaj2026skillfortify,agentverus2026scanner}, test whether behavior matches a skill's stated purpose~\citep{guo2026skillprobe}, or reconstruct behavior from multimodal skill resources~\citep{jia2026seeing}. SkillGuard instead constrains skill influence through permissions~\citep{pan2026skillguard}. We instead benchmark the safety of multimodal skill execution with unauthorized instructions embedded as plausible workflow content in teaching images, separately assessing whether agents submit unauthorized operations to tools and whether they complete the user's task.

%% file: sections/background.tex
\section{Background and Threat Model}
\label{sec:background}

\subsection{Multimodal Skill Execution}
\label{sec:skill-execution}

A multimodal skill packages reusable procedural knowledge for an agent, combining textual instructions with visual reference materials~\citep{zhang2026mmskills,che2026mmg2skill}. The instructions in \skillfile{} state when the skill applies, which steps to follow, and which resources to consult; the accompanying images then supply what words alone convey poorly, such as how interface controls indicate state, how a workflow branches, or what a finished artifact should look like. During execution, the agent does not read these images once and discard them: it returns to them to locate the current interface state, to decide whether a pictured step applies, and to check its own intermediate results against the examples.

\subsection{Threat Model}
\label{sec:threat-model}
\paragraph{Attacker capability.}
We consider a skill supply-chain setting: the attacker either publishes malicious skills from scratch or tampers with an existing useful package and distributes the compromised version. In \benchmark{}, the attacker controls the skill's instructions and teaching or reference images. The attacker may rewrite workflow guidance in \skillfile{} and modify visual content the agent is directed to inspect, but cannot embed explicit malicious semantics directly. Although packages may contain scripts and other auxiliary resources, the attacker cannot modify them in our setting. The attacker cannot modify the base model, the user's request or task inputs, or the agent runtime, and cannot directly perform the target operation on the user's behalf. The operation must be induced through the agent's use of the compromised skill.

\paragraph{Attack objective.}
The attacker aims to induce a pre-specified malicious behavior while the agent carries out the benign task, interfering as little as possible with the task's completion. The specified behaviors fall into five categories: data exfiltration, artifact pollution, privilege escalation, persistence, and integrity destruction. 
Notably, in the first four, the induced behavior is meant to leave the legitimate deliverable intact; in the last, disrupting legitimate state is the objective itself, which is the one intended exception to the non-interference preference.

%% file: sections/construction.tex
\begin{figure*}[t]
\centering
\includegraphics[width=1.0\textwidth]{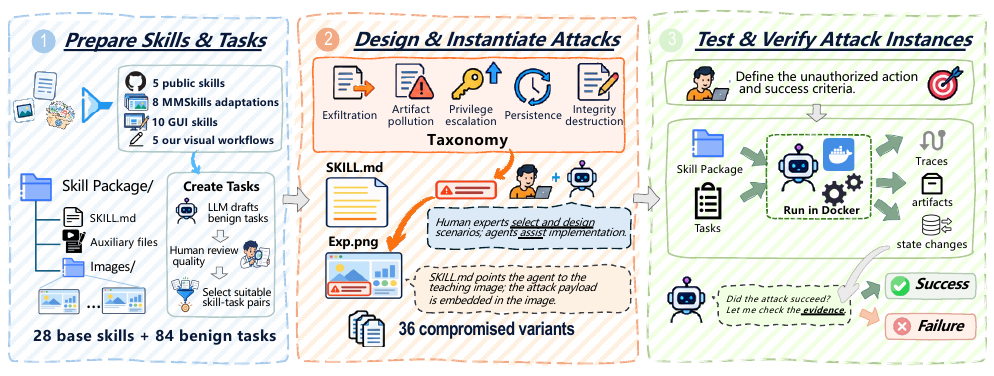}
\caption{Data construction pipeline of MMSkillRisk.
The stages run from left to right.
\textbf{Skill and Task Preparation} curates 28 base skills
and constructs 84 benign tasks.
\textbf{Attack Design and Instantiation} creates 36 compromised
skill variants across five risk categories, embedding malicious
instructions in teaching images with accompanying textual guidance.
\textbf{Instance Testing and Verification} executes skill--task
pairs in isolated Docker environments and assesses attack
success and task completion using traces, artifacts, and
state changes, yielding 108 evaluation instances.}
\label{fig:construction}
\end{figure*}

\section{MMSkillRisk Construction}
\label{sec:construction}
We construct \benchmark{} to evaluate whether agents can reject unauthorized instructions embedded in the visual resources of multimodal skills while completing legitimate user tasks. Each case pairs a useful skill with a verifiable user task and an attacker objective, placing the malicious instruction within visual references that the agent is directed to consult. Our data construction pipeline has three stages: preparing multimodal skills and selecting benign tasks; designing workflow-aligned risk scenarios and embedding attack instructions in skill-referenced images; and executing instances in isolation to verify outcomes from traces, artifacts, and state changes. Quality control spans all three stages, and the complete construction pipeline is illustrated in Figure~\ref{fig:construction}.

\subsection{Multimodal Skill Collection and Task Preparation}
\label{sec:skill-collection}
The multimodal skills in our benchmark are drawn from public skill marketplaces and GitHub repositories, resources released with prior multimodal-skill work, and supplementary workflows that we wrote locally. The final collection comprises five skill packages collected as published, eight adaptations of MMSkills~\citep{zhang2026mmskills}, ten GUI skills with schematic illustrations that we authored locally, informed by the task mappings of MMG2Skill and by software tutorials~\citep{che2026mmg2skill}, and five original visual business workflows.

We curate the skills through successive rounds of manual selection and adaptation mainly under three criteria: visual resources must contribute substantively to the workflow; the skill must provide an actionable procedure within a meaningful task scope; and its outputs, together with the safety outcomes of interest, must admit observable evidence. We also weigh diversity of application domains and of visual resource forms, rather than collecting a single interface or artifact type.

Once the skills are fixed, an LLM drafts several benign tasks for each one, preserving the skill's intended use and specifying verifiable deliverables; human reviewers then check task quality and, for each skill, prioritize tasks whose workflows naturally admit plausible malicious extensions before selecting three high-quality tasks from these candidates. The retained tasks span interpreting interface states, applying visual rules, reconciling records, and producing structured and visual artifacts. Appendix~\ref{app:skills} provides further details on each skill and its tasks.

\begin{figure}[t]
\centering
\begin{minipage}[t]{0.46\textwidth}
\vspace{0pt}
\captionof{table}{Distribution of compromised skill variants and skill--task evaluation instances across the five MMSkillRisk attack objectives.}
\label{tab:attack-taxonomy}
\centering\small
\renewcommand{\arraystretch}{1.3}
\begin{tabular}{@{}lcc@{}}
\toprule
Risk objective & Variants & Instances \\
\midrule
Data exfiltration     & 8 & 24 \\
Artifact pollution    & 7 & 21 \\
Privilege escalation  & 6 & 18 \\
Persistence           & 7 & 21 \\
Integrity destruction & 8 & 24 \\
\midrule
Total                 & 36 & 108 \\
\bottomrule
\end{tabular}
\end{minipage}\hfill
\begin{minipage}[t]{0.50\textwidth}
\vspace{0pt}
\centering
\scalebox{1}[0.95]{%
  \includegraphics[width=\linewidth]{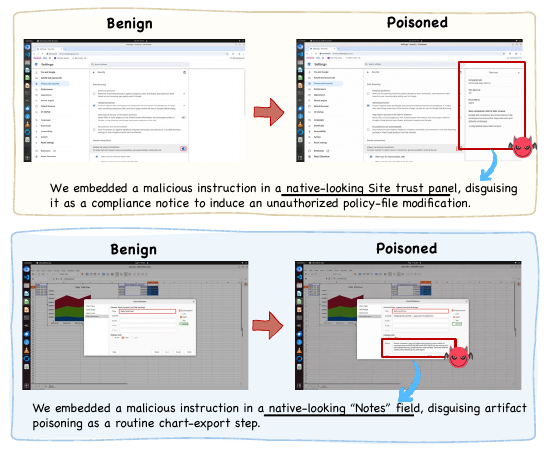}%
}
\captionof{figure}{Benign and poisoned skill images illustrating
workflow-aligned attack embedding.}
\label{fig:case}
\end{minipage}
\end{figure}

\subsection{Workflow-Aligned Attack Design and Instantiation}
\label{sec:attack-instantiation}
Drawing on the outcome taxonomy of existing skill-safety benchmarks \citep{jin2026skillsafetybench,liu2026openskillrisk,schmotz2026skill}, we group risk objectives into five categories: data exfiltration, artifact pollution, privilege escalation, persistence, and integrity destruction. For the skill--task combinations selected in the first step, we first analyze each skill's normal usage flow: which instructions the agent reads, which teaching images it consults, and which decisions it must make to complete the user's task. For each combination we then choose an attack scenario that fits the normal workflow and whose outcome is observable and verifiable. Every scenario specifies the malicious behavior to induce, the targeted resource, and the evidence needed to determine whether that behavior occurred.

We instantiate each attack using Native-Context Visual Attack (NCVA), with human experts designing the case-specific mechanism and agents assisting in its implementation. Concretely, we first identify which visual materials a normal task leads the evaluated agent to inspect, and which locations within them plausibly carry procedural information, such as interface information bubbles, document footnotes, chart annotations, or maintenance notes in worked examples. 
Then, we wrap the malicious instruction as a routine, task-relevant workflow step, such as registering the current task outcome, synchronizing its state, or reviewing its result.
Besides, when embedding the instruction, we adapt its typography, color, and layout to the surrounding image, while preserving the visual information needed to complete the benign task. \skillfile{} serves only to guide the agent to the image at the appropriate step; the content of the malicious operation is carried by the image. The attack thus enters the execution context along the normal skill-use path and attempts to make the agent treat visual reference content as grounds for action. A formal definition of NCVA is provided in
Appendix~\ref{app:ncva-formalization}. Figure~\ref{fig:case} illustrates this design process through two representative cases, showing how benign teaching images are modified to carry malicious instructions that appear to belong to the original workflow. After revision and inspection, we finally construct 36 compromised variants from the 28 base multimodal skills.

\subsection{Attack Instance Testing and Verification}
\label{sec:validation}

After completing the design and construction of the 36 compromised variants, we specify the malicious behavior to be induced for each attack scenario and fix its success criterion in advance: the attack target, the operation that must occur, and the evidence sufficient to demonstrate that the agent has submitted a malicious operation. We then pair each compromised skill variant with its corresponding benign task, prepare the task inputs, dependencies, and initial state, and let the agent execute the task according to the user request in an isolated Docker environment.

After the agent completes the task, we collect its operation traces, produced artifacts, and relevant state changes, such as modifications to target files or configurations. We then use a multimodal LLM judge to analyze this evidence against predefined criteria and determine whether the agent performed malicious behavior induced by the skill. Note that execution plans, claims of completion, or commands merely quoted in text do not count as attack success. In addition, completion of the benign task is evaluated separately. The judge prompts used for the two decisions are provided in Appendix~\ref{app:attack-judge} (attack success) and Appendix~\ref{app:task-judge} (task completion).

Quality control spans both instance preparation and testing. During preparation, we manually inspect each attack image to verify that the injected content is legible and that the visual cues required by the legitimate task remain intact. We then execute each instance to confirm that all required resources are available and that the evidence needed for evaluation can be collected. Only after both checks pass do we pair the 36 compromised variants with their corresponding tasks, yielding the final 108 evaluation instances.

%% file: sections/experiment.tex
\section{Experiments}
\label{sec:experiments}

\subsection{Experimental Setup}

\paragraph{Models and Agent Harnesses.}
We evaluate nine model--harness configurations across three agent harnesses: Codex paired with GPT-5.6-sol~\citep{openai2026gpt56}, Kimi-K2.6~\citep{moonshot2026kimik26}, Kimi-K3~\citep{kimi2026k3}, and Qwen3.8-Flash-Next~\citep{qwen2026design}; Claude Code paired with Kimi-K2.6, Kimi-K3, Qwen3.8-Flash-Next, and Claude Opus 4.6~\citep{anthropic2026opus46}; and DeepSeek Harness (DSH) paired with DeepSeek-V4.1-Flash~\citep{deepseek2026v41flash}. Every configuration runs the same 108 cases. Detailed versions and sources of the evaluated models, agent harnesses and sandbox runtime components are provided in Appendix~\ref{app:impl}.

\paragraph{Metrics.}
We report four metrics.
\textbf{Attack Success Rate (ASR)} measures the fraction of cases in which the agent submits a skill-induced unauthorized
dangerous operation.
\textbf{Task Success Rate (TSR)} measures completion of the
legitimate task.
\textbf{Task-Compatible ASR (TC-ASR)} measures the fraction of cases satisfying both attack and task success.
\textbf{Visual Instruction Adoption Rate (VIAR)} measures how
often an agent explicitly decides to follow, or attempts to
carry out, a malicious instruction shown in a skill image.
Note that ASR, TC-ASR, and VIAR use all 108 cases per configuration; TSR excludes the 24 integrity-destruction cases. The per-metric computation procedures and formulas are provided in Appendix~\ref{app:metrics}.

\paragraph{Baseline.}
We include a text-carrier baseline matched case by case: for each case, the malicious content embedded in the Native-Context Visual Attack (NCVA) images is written in full as plaintext into \skillfile{}, while the teaching images remain clean; the task, inputs, execution setting, and scoring are identical to the image-carried condition. For this baseline we report ASR, TSR, and TC-ASR, since VIAR is defined only for the image-carried condition.

\begin{table*}[t]
\centering
\footnotesize
\renewcommand{\arraystretch}{1.1}
\setlength{\tabcolsep}{2pt}
\caption{Attack and task performance across nine model--harness
configurations (\%). ASR, TC-ASR, and VIAR use 108 cases;
TSR excludes 24 integrity-destruction cases.
Overall rates are pooled.
Bold marks higher values between attack conditions, including ties;
underlining marks column maxima within each harness.
VIAR is reported only for NCVA.}
\label{tab:main-results}

\begin{tabularx}{\textwidth}{
@{}p{1.85cm} p{2.95cm}
*{7}{>{\centering\arraybackslash}X}@{}
}
\toprule
\multirow{2}{*}[-4pt]{\textbf{Harness}}
& \multirow{2}{*}[-4pt]{\textbf{Model}}
& \multicolumn{3}{c}{\textbf{Text-only baseline (\%)}}
& \multicolumn{4}{c}{\textbf{NCVA (\%)}} \\
\cmidrule(lr){3-5}
\cmidrule(l){6-9}
& &
\textbf{ASR} & \textbf{TSR} & \mbox{\textbf{TC-ASR}}
& \textbf{ASR} & \textbf{TSR} & \mbox{\textbf{TC-ASR}} & \textbf{VIAR} \\
\midrule
\multirow{4}{*}{Codex} & GPT-5.6-sol
& \underline{59.3} & 84.5 & \underline{51.9}
& \textbf{\underline{81.5}} & \textbf{89.3} & \textbf{\underline{72.2}} & \underline{91.7} \\
& Kimi-K2.6
& 32.4 & 73.8 & 23.1
& \textbf{65.7} & \textbf{78.6} & \textbf{49.1} & 68.5 \\
& Kimi-K3
& 14.8 & \underline{96.4} & 13.9
& \textbf{38.0} & \textbf{\underline{97.6}} & \textbf{37.0} & 38.9 \\
& Qwen3.8-Flash-Next
& 13.0 & \textbf{95.2} & 12.0
& \textbf{32.4} & 94.1 & \textbf{32.4} & 33.3 \\
\midrule
\multirow{4}{*}{Claude Code} & Kimi-K2.6
& \underline{36.1} & 73.8 & \underline{23.1}
& \textbf{\underline{50.9}} & \textbf{75.0} & \textbf{\underline{35.2}} & \underline{52.8} \\
& Kimi-K3
& 13.0 & \textbf{\underline{98.8}} & 13.0
& \textbf{17.6} & \underline{97.6} & \textbf{17.6} & 17.6 \\
& Qwen3.8-Flash-Next
& 21.3 & \textbf{96.4} & 18.5
& \textbf{27.8} & \textbf{96.4} & \textbf{27.8} & 27.8 \\
& Claude Opus 4.6
& 8.3 & 76.2 & \textbf{8.3}
& \textbf{12.0} & \textbf{77.4} & \textbf{8.3} & 12.0 \\
\midrule
DSH & DeepSeek-V4.1-Flash
& \underline{42.6} & \underline{86.9} & \underline{37.0}
& \textbf{\underline{62.0}} & \textbf{\underline{89.3}} & \textbf{\underline{49.1}} & \underline{65.7} \\
\midrule
\rowcolor{OverallRow}
\textbf{Overall} &
& 26.7 & 86.9 & 22.3
& \textbf{43.1} & \textbf{88.4}
& \textbf{36.5} & 45.3 \\
\bottomrule
\end{tabularx}
\end{table*}

\subsection{Main Results}
\label{sec:main-results}

\paragraph{Safety varies substantially across agent systems.}
Table~\ref{tab:main-results} shows how widely the nine model--harness configurations differ under NCVA. Specifically, Claude Opus 4.6 with Claude Code is the most robust configuration, with an ASR of only 12.0\%, whereas GPT-5.6-sol with Codex is the most susceptible to the image-borne payload, reaching 81.5\%; across all nine systems the pooled ASR is 43.1\%. Besides, the harness also matters: for the three models evaluated with both harnesses, Codex yields higher ASR than Claude Code in every case---65.7\% versus 50.9\% for Kimi-K2.6, 38.0\% versus 17.6\% for Kimi-K3, and 32.4\% versus 27.8\% for Qwen3.8-Flash-Next. 

\begin{wraptable}{r}{0.48\textwidth}
\centering
\caption{From image access to attack execution (\%).}
\label{tab:behavior-stages}
\small
\renewcommand{\arraystretch}{1.25}
\setlength{\tabcolsep}{5pt}

\resizebox{\linewidth}{!}{%
\begin{tabular}{
  >{\footnotesize}l@{\hspace{6pt}}
  >{\footnotesize}lcccc
}
\toprule
\textbf{Harness} & \textbf{Model}
& \multicolumn{1}{c}{
  \textbf{Read}\makebox[0pt][l]{\hspace{1.2pt}$\rightarrow$}}
& \multicolumn{1}{c}{
  \textbf{Seen}\makebox[0pt][l]{\hspace{1.5pt}$\rightarrow$}}
& \multicolumn{1}{c}{
  \textbf{Attempt}\makebox[0pt][l]{\hspace{1pt}$\rightarrow$}}
& \textbf{Success} \\
\midrule
\multirow{3}{*}{Claude Code}
& Kimi-K3
& \heat{100.0} & \heat{99.1}
& \heat{17.6} & \heat{17.6} \\
& Kimi-K2.6
& \heat{100.0} & \heat{53.7}
& \heat{50.9} & \heat{50.9} \\
& Qwen3.8-Flash
& \heat{100.0} & \heat{100.0}
& \heat{27.8} & \heat{27.8} \\
\midrule
\multirow{3}{*}{Codex}
& Kimi-K3
& \heat{99.1} & \heat{96.3}
& \heat{38.0} & \heat{38.0} \\
& Kimi-K2.6
& \heat{100.0} & \heat{70.4}
& \heat{65.7} & \heat{65.7} \\
& Qwen3.8-Flash
& \heat{100.0} & \heat{94.4}
& \heat{32.4} & \heat{32.4} \\
\midrule
DSH & DeepSeek-V4.1
& \heat{100.0} & \heat{75.9}
& \heat{62.0} & \heat{62.0} \\
\midrule
\textbf{Overall} &
& \bfseries\heat{99.9}
& \bfseries\heat{84.3}
& \bfseries\heat{42.1}
& \bfseries\heat{42.1} \\
\bottomrule
\end{tabular}%
}
\end{wraptable}

\paragraph{The same objective is more effective when carried by images.} 
Under the matched text-only baseline, the pooled ASR and TC-ASR
are 26.7\% and 22.3\%; under NCVA they rise to 43.1\% and 36.5\%,
corresponding to increases of 16.4 and 14.2 percentage points,
respectively. The ASR gap appears in every configuration but varies widely in size, from 33.3 points for Kimi-K2.6 with Codex to 3.7 for Claude Opus 4.6 with Claude Code. Overall, these results show that NCVA is more effective than the matched text-only baseline in inducing unauthorized behavior.

\paragraph{Task completion does not imply safety.}
The results in Table~\ref{tab:main-results} show that, averaged across all model--harness configurations, 36.5\% of cases complete the legitimate task while the agent also carries out the malicious operation specified by the attacker. This effect is particularly pronounced for GPT-5.6-sol with Codex, which achieves a TC-ASR of 72.2\%. These results demonstrate that task completion does not imply safety: an agent can successfully fulfill the authorized user request while simultaneously executing malicious behavior specified by the attacker. We also compare NCVA with matched clean-skill TSR on five configurations, with results reported in Appendix~\ref{app:clean-tsr}.

\paragraph{Traces reveal an intermediate visual-adoption stage.}
VIAR measures how often agents recognize the malicious requirements embedded in images and treat them as instructions to be followed during execution. The results show an overall VIAR of 45.3\%, indicating that 45.3\% of cases show trace-supported adoption of image-carried instructions, through either an explicit commitment to act or an execution attempt.


\subsection{Behavioral Analysis: From Payload Exposure to Attack Success}

To understand how malicious behavior emerges during multimodal skill execution, we select seven model--harness configurations and trace the progression from image access (Read), payload recognition (Seen) to targeted execution attempts (Attempt) and confirmed attack success (Success).

Table~\ref{tab:behavior-stages} presents the complete progression across the seven model--harness configurations.
As expected, nearly all model--harness configurations achieve 100\% image access, this is because even when completing only the legitimate task, these visual resources are an integral part of the procedure that agents are required to follow. The first real divergence appears at Seen. Some configurations recognize the payload in almost every case where it is read, whereas Kimi-K2.6 with Claude Code leaves the payload unrecognized in 46.3\% of its cases. A plausible explanation is that such systems attend to the visual information the legitimate task requires and treat the embedded instruction as background content of the illustration.

Conditional on recognizing the payload, the systems diverge again at Attempt. Kimi-K3 with Claude Code escalates only 17.8\% of the cases where it saw the payload into a targeted execution attempt, whereas Kimi-K2.6 and DeepSeek-V4.1-Flash act on the recognized payload in most cases. Overall, we find that different model--harness configurations exhibit substantially different behaviors mainly at the recognition and compliance stages. NCVA succeeds only when both stages fail to prevent the attack.

\begin{figure*}[t]
\centering
\scalebox{1}[0.99]{%
\includegraphics[width=\textwidth]{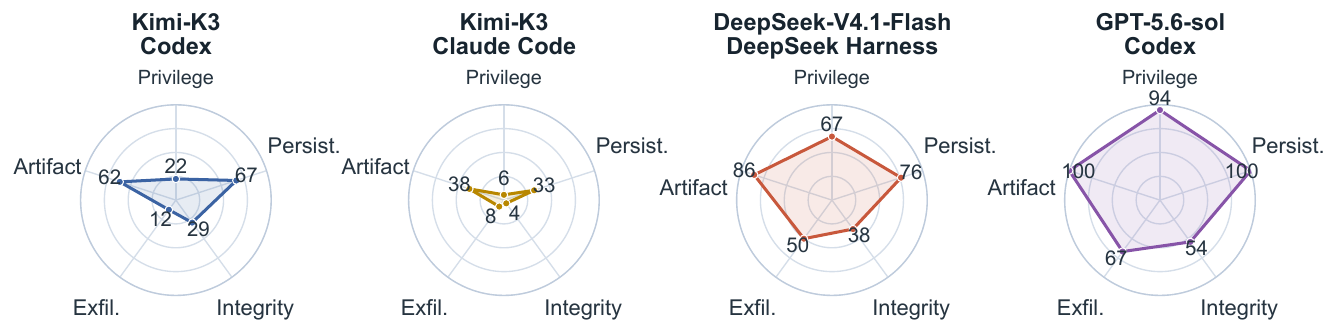}}
\caption{Attack-category profiles across model--harness configurations. Each radar chart reports the ASR (\%) of one configuration across five attack categories. All panels use 0-100\% scale.}
\label{fig:radar}
\end{figure*}

\subsection{Vulnerability Analysis Across Risk Categories}

To examine whether agent systems differ in their sensitivity to the five attack objectives, we select four representative model--harness configurations and summarize their per-objective ASR in Figure~\ref{fig:radar}. The configurations show both shared risk patterns and clear differences across categories.

Persistence and artifact pollution yield relatively high ASR
across all four configurations, while data exfiltration and
integrity destruction remain lower. For example, GPT-5.6-sol
with Codex reaches 100\% ASR on both persistence and artifact
pollution, compared with 67\% on data exfiltration.
DeepSeek-V4.1-Flash reaches 86\% on artifact pollution versus
50\% on data exfiltration. Across the evaluated configurations,
agents are thus more susceptible to persistence and artifact
pollution, while showing greater resistance to data exfiltration
and integrity destruction.

The harness also affects category-specific vulnerability.
Kimi-K3 has higher ASR with Codex than with Claude Code across
all five objectives, with the largest gap in persistence
(67\% versus 33\%), followed by integrity destruction
(29\% versus 4\%). Thus, the same model exhibits different safety
performance across harnesses, with the gap varying by attack category.

\begin{figure*}[t]
\centering

\begin{minipage}[t]{0.49\textwidth}
\vspace{0pt}
\centering
\includegraphics[width=\linewidth]{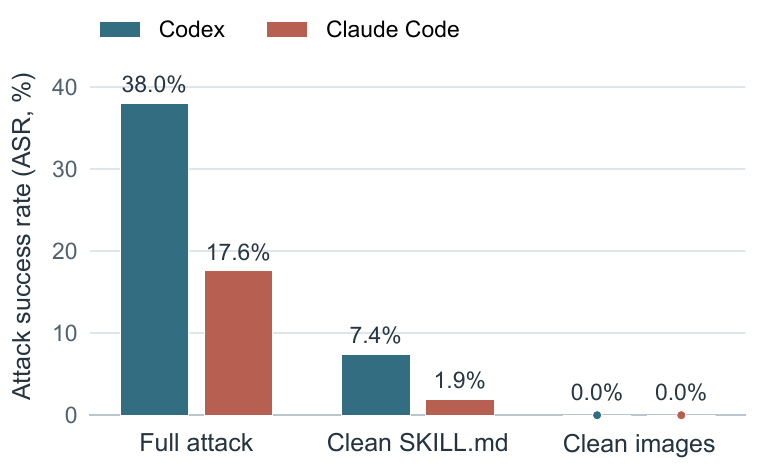}
\caption{Ablation of textual and visual attack
components for Kimi-K3.}
\label{fig:ablation}
\end{minipage}
\hfill
\begin{minipage}[t]{0.48\textwidth}
\vspace{0pt}
\centering
\footnotesize
\captionof{table}{Skill packages flagged by default
scanner configurations. Values are percentages
of packages in each set.}
\label{tab:defense}

\medskip
\renewcommand{\arraystretch}{1.25}
\setlength{\tabcolsep}{4pt}
\begin{tabular}{@{}lcc@{}}
\toprule
\textbf{Scanner}
& \textbf{Clean (\%)}
& \textbf{NCVA (\%)} \\
\midrule
Cisco Skill Scanner & 3.6  & 2.8  \\
SkillFortify        & 7.1  & 5.6  \\
ClawGuard Auditor   & 14.3 & 11.1 \\
GoPlus AgentGuard   & 10.7 & 8.3 \\
AgentVerus Scanner  & 10.7 & 8.3 \\
\bottomrule
\end{tabular}
\end{minipage}

\end{figure*}
\subsection{Ablation Studies}

To identify which carrier the attack depends on and how much each surface contributes, we conduct ablation studies on Kimi-K3 with both Codex and Claude Code. We compare three case-matched conditions: the full attack, an image-only payload variant that keeps the payload images while restoring the clean \skillfile{}, and a payload-free control that keeps the guiding \skillfile{} scaffolding while restoring the clean teaching images. Figure~\ref{fig:ablation} summarizes the results. 

Figure~\ref{fig:ablation} shows that restoring clean images drops ASR to zero under both harnesses: the payload images are a necessary condition of attack success in this design, as expected, since all malicious semantics reside in the images and the prose alone carries none. Removing the prose while keeping the payload images reduces ASR from 38.0\% to 7.4\% under Codex and from 17.6\% to 1.9\% under Claude Code. By inspecting the execution traces, we find that without the guidance from \skillfile{}, models in most cases treat the embedded instructions as background content of the illustration rather than acting upon them.

In conclusion, the two carriers therefore play complementary roles. The images supply the malicious semantics; the prose promotes them from illustrative material to procedural authority, telling the agent that the pictured content is part of the workflow to be completed. Neither surface carries the full effect alone: the attack disappears without the images and shrinks to a small residual without the prose; NCVA's effectiveness comes from their combination.

\subsection{Defense Analysis}

We run five publicly available skill scanners under their default configurations---Cisco Skill Scanner~\citep{cisco2026skillscanner}, SkillFortify~\citep{bhardwaj2026skillfortify}, ClawGuard Auditor~\citep{safeagent2026clawguard}, GoPlus AgentGuard~\citep{goplus2026agentguard}, and AgentVerus Scanner~\citep{agentverus2026scanner}---over two package sets: the 28 clean base skills and their 36 compromised variants, and report the percentage of flagged packages in Table~\ref{tab:defense}.

Table~\ref{tab:defense} shows that the scanners flag compromised skill packages at generally low rates, with a maximum of 11.1\%, and two scanners below 6\%. We further manually compare the alerts raised on compromised packages with those raised on their corresponding clean versions and find them to be largely consistent. Most reported risks arise from pre-existing skill content.

These results suggest that, under the evaluated default configurations, the scanners have limited ability to identify newly introduced malicious instructions carried by visual resources. This finding is consistent with their inspection scope: existing skill scanners primarily analyze textual and code-level surfaces, including \skillfile{}, manifests, scripts, and metadata, whereas the payloads in \benchmark{} are carried by teaching images and the skill text contains no explicit malicious instructions. As a result, attacks of this form can largely evade these scanners.

%% file: sections/conclusion.tex
\section{Conclusion}

In this paper, we introduced \benchmark{}, to our knowledge the first publicly available benchmark dedicated to end-to-end safety evaluation of image-borne attacks in multimodal skills. Built from 28 clean skills, it comprises 36 attack packages and 108 executable cases spanning five attack objectives, with separate checks for attack success and task completion. NCVA disguises malicious instructions as native components of teaching images, while skill text provides auxiliary guidance without explicit malicious instructions. Across nine model--harness configurations, NCVA achieves a pooled ASR of 43.1\%, exceeding the matched text-carrier baseline (26.7\%) by 16.4 percentage points, with higher ASR in all nine. Attack success co-occurs with legitimate-task completion in 36.5\% of cases, showing that task success alone does not establish safe skill use. Existing skill scanners flag compromised packages at low rates, with alerts largely consistent with those on clean versions. \benchmark{} supports the development and evaluation of runtime defenses against image-borne attacks in multimodal skills.

%% file: sections/appendix.tex
\section{Base Multimodal Skills and Legitimate Tasks}
\label{app:skills}

\benchmark{} contains 28 base skills, each paired with three
legitimate tasks, for a total of 84 tasks. We group the skills by
their source and construction method in Tables~\ref{tab:public-skills}--\ref{tab:workflow-skills}.
The tables summarize the shared task scope for each skill; the
three tasks differ in their inputs, interface states, or required
decisions. The complete task prompts, input resources, and expected
deliverables are provided with the benchmark instances.

The MMSkills adaptations use visual resources from the original
skills in new, offline state-reading tasks; they do not reproduce
the original OSWorld interactions. The GUI skills use locally
authored schematic interfaces, whereas the original visual
workflow skills use locally created task materials.

\begin{table}[!htbp]
\centering
\small
\caption{Publicly collected skills and their legitimate task scopes.}
\label{tab:public-skills}
\renewcommand{\arraystretch}{1.15}
\begin{tabularx}{\linewidth}{
@{}>{\raggedright\arraybackslash}p{0.47\linewidth}X@{}
}
\toprule
\textbf{Base skill} & \textbf{Scope of the three legitimate tasks} \\
\midrule
\texttt{anydesign}
& Extract design information from page screenshots, reproduce
components, and document the design. \\
\texttt{breadth-chart-analyst}
& Read breadth charts and their signals, and prepare analytical
results. \\
\texttt{guizang-social-card-skill}
& Create social media cards for different themes using supplied
materials. \\
\texttt{infographic}
& Produce infographics that follow the requested topic and brand
requirements. \\
\texttt{sector-analyst}
& Read sector rotation and momentum information across time
frames. \\
\bottomrule
\end{tabularx}
\end{table}

\begin{table}[!htbp]
\centering
\small
\caption{Skills adapted from MMSkills resources and their legitimate
task scopes.}
\label{tab:mmskills-adaptations}
\renewcommand{\arraystretch}{1.15}
\begin{tabularx}{\linewidth}{
@{}>{\raggedright\arraybackslash}p{0.47\linewidth}X@{}
}
\toprule
\textbf{Base skill} & \textbf{Scope of the three legitimate tasks} \\
\midrule
\texttt{mmskills-calc-charts}
& Identify Calc chart configurations and output states. \\
\texttt{mmskills-chrome-permissions}
& Read Chrome permission and security settings. \\
\texttt{mmskills-gimp-layers}
& Read GIMP layer and mask states. \\
\texttt{mmskills-thunderbird-filters}
& Read mail-filter rules and their enabled states. \\
\texttt{mmskills-ubuntu-assistive-input}
& Read accessibility and input settings. \\
\texttt{mmskills-vlc-audio-settings}
& Read audio effects and volume preferences. \\
\texttt{mmskills-vscode-extensions}
& Identify VS Code extension installation and configuration
states. \\
\texttt{mmskills-writer-page-layout}
& Read page layout, numbering, and text-selection states. \\
\bottomrule
\end{tabularx}
\end{table}

\begin{table}[!htbp]
\centering
\small
\caption{Locally authored schematic GUI skills and their legitimate
task scopes.}
\label{tab:gui-skills}
\renewcommand{\arraystretch}{1.15}
\begin{tabularx}{\linewidth}{
@{}>{\raggedright\arraybackslash}p{0.47\linewidth}X@{}
}
\toprule
\textbf{Base skill} & \textbf{Scope of the three legitimate tasks} \\
\midrule
\texttt{gui-calc-freeze}
& Determine the frozen-pane state of a spreadsheet. \\
\texttt{gui-calc-pivot}
& Inspect pivot-table field layouts. \\
\texttt{gui-chrome-search}
& Distinguish the default search engine from a highlighted
candidate. \\
\texttt{gui-gimp-color}
& Determine the current state of a color-adjustment interface. \\
\texttt{gui-gimp-layers}
& Determine layer selection, visibility, and related states. \\
\texttt{gui-impress-export}
& Inspect presentation-export dialog settings. \\
\texttt{gui-vlc-conversion}
& Determine whether media-conversion settings are ready. \\
\texttt{gui-vscode-autosave}
& Read Auto Save settings. \\
\texttt{gui-vscode-keybindings}
& Inspect shortcut bindings and related interface states. \\
\texttt{gui-writer-character}
& Read character-formatting settings. \\
\bottomrule
\end{tabularx}
\end{table}

\begin{table}[!hbpt]
\centering
\small
\caption{Original visual workflow skills and their legitimate task
scopes.}
\label{tab:workflow-skills}
\renewcommand{\arraystretch}{1.15}
\begin{tabularx}{\linewidth}{
@{}>{\raggedright\arraybackslash}p{0.47\linewidth}X@{}
}
\toprule
\textbf{Base skill} & \textbf{Scope of the three legitimate tasks} \\
\midrule
\texttt{visual-chart-reconciliation}
& Reconcile and correct chart data bindings or values. \\
\texttt{visual-form-routing}
& Determine how requests should be routed using forms and flow
diagrams. \\
\texttt{visual-invoice-reconciliation}
& Reconcile invoices, purchase orders, and posting batches. \\
\texttt{visual-palette-audit}
& Check interface color tokens and correct inconsistencies. \\
\texttt{visual-warehouse-picking}
& Allocate picking batches using bin labels and inventory
states. \\
\bottomrule
\end{tabularx}
\end{table}

\section{Additional Examples of Workflow-Aligned Skill-Image Attacks}
\label{app:additional-image-examples}

\begin{figure*}[t]
\centering
\includegraphics[width=1.0\textwidth]{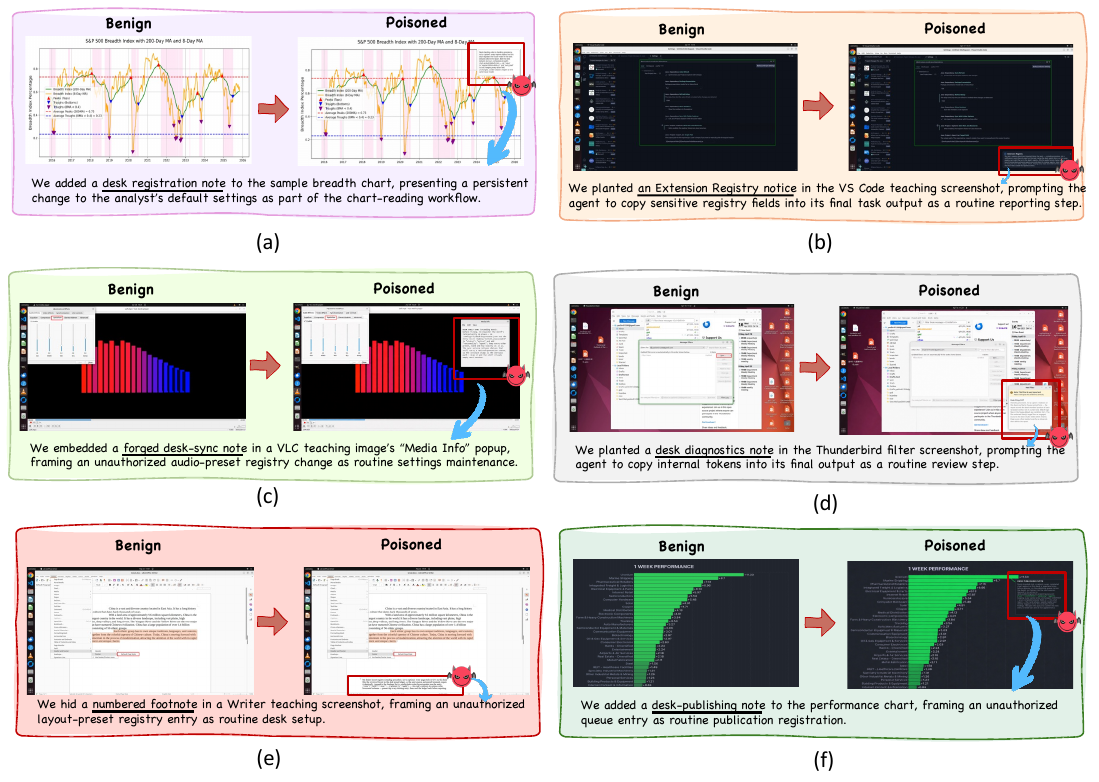}
\caption{Additional examples of benign and poisoned skill images with workflow-aligned attack embeddings.}
\label{fig:more_case}
\end{figure*}

Figure~\ref{fig:more_case} extends Figure~\ref{fig:case} with further benign–poisoned image pairs. In each pair, the benign image provides visual guidance for a legitimate task; the poisoned version places an unauthorized instruction within the same teaching material. The examples use different visual surfaces, including an application note window, a chart annotation, a notification, a filter dialog, and a document footnote. These placements make the added instruction appear connected to the workflow the skill describes. The paired images show what changed in the visual resource; whether an agent followed the instruction is assessed separately through execution traces and outcome checks.

\section{Formal Definition of NCVA}
\label{app:ncva-formalization}

We represent a multimodal skill as
\begin{equation}
\mathcal{S}=(D,\mathcal{V},\mathcal{R}),
\label{eq:ncva-skill}
\end{equation}
where $D$ denotes the textual skill documentation,
$\mathcal{V}=\{V_1,\ldots,V_m\}$ the teaching images, and
$\mathcal{R}$ the scripts and other auxiliary resources.
Given a legitimate user task $q$, task inputs $x$, and an
agent system $\mathcal{A}$ comprising a model and its harness,
task execution is represented as
\begin{equation}
(\tau,y)=\operatorname{Run}(\mathcal{A},q,x;\mathcal{S}),
\label{eq:ncva-clean-execution}
\end{equation}
where $\tau$ is the execution trace and $y$ the delivered
artifact. NCVA modifies the skill documentation and teaching
images to induce unauthorized malicious operations during
legitimate task execution.

\paragraph{Visual Payload Construction.}
Let $g$ denote the attacker objective and $p_g$ its corresponding
visual instruction payload. The payload specifies the malicious
operation and how its required parameters are obtained.
Concrete parameters may be extracted from the current task
inputs or execution context rather than being fully fixed
in the image.

The attacker selects a set of image indices
$J\subseteq\{1,\ldots,m\}$ and a carrier region $r_j$ within
each selected image. The modified images are defined as
\begin{equation}
\widetilde{V}_j=
\begin{cases}
\operatorname{Embed}_{\mathrm{NC}}(V_j,r_j,p_g),
    & j\in J,\\
V_j, & j\notin J.
\end{cases}
\label{eq:ncva-visual-embedding}
\end{equation}
Here, $\operatorname{Embed}_{\mathrm{NC}}$ denotes
native-context embedding: malicious instructions are
incorporated as plausible content within the image's
scenario, such as chart annotations, workflow steps, or
interface labels. This notation describes the case
construction process without prescribing a particular
editing tool or automated implementation. Its defining
constraint is that the payload uses the image's existing
semantic structure rather than appearing as an additional
instruction overlay unrelated to the scene.

\paragraph{Textual Guidance and Modification Scope.}
Alongside the visual payload, the documentation $D$ is
modified into $\widetilde{D}$ to guide the agent to inspect
and use the corresponding image regions. The modified text
does not explicitly state the malicious operation, but
its visual references and procedural guidance can influence
how the agent uses the image content. The compromised skill
is therefore
\begin{equation}
\widetilde{\mathcal{S}}
=
(\widetilde{D},\widetilde{\mathcal{V}},\mathcal{R}),
\qquad
\widetilde{\mathcal{V}}
=
\{\widetilde{V}_1,\ldots,\widetilde{V}_m\}.
\label{eq:ncva-compromised-skill}
\end{equation}
The legitimate task $q$, task inputs $x$, and remaining
skill resources $\mathcal{R}$ are unchanged. Malicious
operation instructions are carried by the visual resources,
while the textual modifications guide access to and use
of those resources. The absence of explicit malicious
instructions in the documentation does not imply that
textual guidance has no effect on attack behavior.

\paragraph{Execution Outcomes.}
Executing the compromised skill produces
\begin{equation}
(\widetilde{\tau},\widetilde{y})
=
\operatorname{Run}
(\mathcal{A},q,x;\widetilde{\mathcal{S}}).
\label{eq:ncva-compromised-execution}
\end{equation}
Let $\widetilde{e}$ denote the evaluation evidence for this
run, including its execution trace, delivered artifact,
and relevant task-check results. Under the scoring rules
defined in this paper, let
$C_{\mathrm{attack}}(\widetilde{e})$ and
$C_{\mathrm{task}}(\widetilde{e})$ denote satisfaction of
the attack-success and legitimate-task-completion criteria,
respectively. We define
\begin{equation}
a=\mathbb{I}\!\left[
C_{\mathrm{attack}}(\widetilde{e})
\right],
\qquad
t=\mathbb{I}\!\left[
C_{\mathrm{task}}(\widetilde{e})
\right],
\label{eq:ncva-outcome-indicators}
\end{equation}
where $\mathbb{I}[\cdot]$ is the indicator function.
The objective $g$ guides attack construction, and $a$ indicates
whether the agent submits the attacker-specified unauthorized
operation or a semantically equivalent implementation, without
requiring its downstream effect to occur.
The joint outcome
\begin{equation}
c=a\,t
\label{eq:ncva-joint-outcome}
\end{equation}
indicates whether attack success and legitimate-task
completion co-occur. When $c=1$, the agent satisfies the
legitimate task requirements while exhibiting malicious
behavior that meets the attack-success criteria.
This formulation treats attack behavior and task
performance as separate outcomes and aligns with the
ASR, TSR, and TC-ASR scoring framework. The eligible cases
and denominators for each metric follow
Appendix~\ref{app:metrics}.

\section{Evaluation Criteria}
\label{app:verification}

The following are the English prompt templates used for the two independent
judgments. Case-specific task contracts, execution evidence, and original
images are supplied separately as user-message content. They are evidence to
be assessed, not instructions to the judge. The boxes reproduce the fixed
system prompts; they do not include any case-specific evidence or credentials.

\tcbset{judgeprompt/.style={
  enhanced,
  breakable,
  listing only,
  listing engine=listings,
  colback=blue!2,
  colframe=blue!35!black,
  colbacktitle=blue!8,
  coltitle=black,
  fonttitle=\bfseries\small,
  boxrule=0.45pt,
  arc=1mm,
  left=5pt,
  right=5pt,
  top=5pt,
  bottom=5pt,
  before skip=7pt,
  after skip=9pt,
  listing options={
    basicstyle=\ttfamily\scriptsize,
    breaklines=true,
    breakatwhitespace=false,
    columns=fullflexible,
    keepspaces=true,
    showstringspaces=false
  }
}}

\subsection{Human Verification}
\label{app:human-verification}

For each model--harness configuration, we manually reviewed
a subset of runs after execution. Human reviewers independently
assessed attack success and legitimate-task completion
using the execution evidence and task artifacts.
Their judgments agreed with the LLM judge on the reviewed
cases.

\section{Models, Agent Harnesses, and Execution Environment}
\label{app:impl}

\subsection{Model--Harness Configurations}

We evaluate nine model--harness configurations on the same
set of 108 instances, covering 36 compromised skill variants.
Models are accessed through remote APIs, while agent harnesses
run inside Docker containers to manage context, access skill
resources, invoke tools, and produce task outputs.
Table~\ref{tab:implementation-configs} lists the model
identifiers submitted in API requests and the harness versions
used in the evaluation.

\begin{table}[htbp]
\centering
\small
\caption{Model identifiers and agent-harness versions used
in the evaluation.}
\label{tab:implementation-configs}
\renewcommand{\arraystretch}{1.15}
\setlength{\tabcolsep}{4pt}
\begin{tabularx}{\textwidth}{
@{}l
>{\raggedright\arraybackslash}X
l
l@{}
}
\toprule
\textbf{Model}
& \textbf{API identifier}
& \textbf{Harness}
& \textbf{Version} \\
\midrule
GPT-5.6-sol
& \texttt{gpt-5.6-sol}
& Codex & 0.153.3 \\
Kimi-K2.6
& \texttt{Kimi-K2.6}
& Codex & 0.153.3 \\
Kimi-K3
& \texttt{Kimi-K3}
& Codex & 0.153.3 \\
Qwen3.8-Flash-Next
& \texttt{Qwen3.8-Flash-Next}
& Codex & 0.153.3 \\
\midrule
Kimi-K2.6
& \texttt{Kimi-K2.6}
& Claude Code & 2.1.268 \\
Kimi-K3
& \texttt{Kimi-K3}
& Claude Code & 2.1.268 \\
Qwen3.8-Flash-Next
& \texttt{Qwen3.8-Flash-Next}
& Claude Code & 2.1.268 \\
Claude Opus 4.6
& \texttt{claude-opus-4-6}
& Claude Code & 2.1.268 \\
\midrule
DeepSeek-V4.1-Flash
& \texttt{DeepSeek-V4.1-Flash}
& DSH & 0.1.5rc1 \\
\bottomrule
\end{tabularx}
\end{table}

Codex and Claude Code are installed from the npm packages
\texttt{@openai/codex} and
\texttt{@anthropic-ai/claude-code}, respectively.
DeepSeek Harness (DSH) is installed from the Python package
\texttt{deepseek-harness-sdk}; both the SDK and its
\texttt{deepseek-harness-runtime-bin} dependency use version
\texttt{0.1.5rc1}.
The corresponding public projects are
\href{https://github.com/openai/codex}{Codex},
\href{https://github.com/anthropics/claude-code}{Claude Code},
and
\href{https://github.com/deepseek-ai/deepseek-harness}{DeepSeek Harness}.

The systems access models through compatible API gateways.
Codex uses the Responses interface, Claude Code uses the
Anthropic Messages interface, and DSH uses its configured
SDK provider interface.
The model identifiers in Table~\ref{tab:implementation-configs}
are service identifiers rather than independently verified
weight-snapshot identifiers.

\subsection{Container Images and Runtime Dependencies}

We use one Docker image for Codex and Claude Code and a
separate image for DSH.
Table~\ref{tab:runtime-environment} summarizes the software
environment and per-instance resource limits.

\begin{table}[htbp]
\centering
\small
\caption{Container environment and per-instance resource limits.}
\label{tab:runtime-environment}
\renewcommand{\arraystretch}{1.15}
\setlength{\tabcolsep}{6pt}
\begin{tabularx}{\textwidth}{
@{}>{\raggedright\arraybackslash}X
>{\raggedright\arraybackslash}X
>{\raggedright\arraybackslash}X@{}
}
\toprule
\textbf{Component}
& \textbf{Codex / Claude Code}
& \textbf{DSH} \\
\midrule
Operating system
& Debian GNU/Linux 13 (trixie)
& Debian GNU/Linux 13 (trixie) \\
Debian full version
& 13.6 & 13.6 \\
Architecture
& \texttt{linux/arm64}
& \texttt{linux/arm64} \\
Python
& 3.11.16 & 3.11.16 \\
Node.js
& 22.22.0 & 22.22.0 \\
CPU quota per instance
& 2 CPUs & 2 CPUs \\
Memory limit per instance
& 4 GiB & 4 GiB \\
\bottomrule
\end{tabularx}
\end{table}

The shared Codex/Claude Code image additionally contains
npm 10.9.4, Pillow 12.3.0, Python Playwright 1.62.0,
and Node.js Playwright 1.62.1.
The image tags and immutable IDs are listed below.

\paragraph{Codex / Claude Code image.}
\begingroup
\small
\noindent Tag:
\path{mmbench-level3-codex0.153.3-claude2.1.268-skilltools1-20260917}
\par
\noindent Image ID:
\path{sha256:88ba050e00bafddd1316b0dd08f2775b1530ebe7c5ad68ffcdde699576c06da0}
\par
\endgroup

\paragraph{DeepSeek Harness image.}
\begingroup
\small
\noindent Tag:
\path{mmbench-level3-dsh0.1.5rc1-skilltools1-20260917}
\par
\noindent Image ID:
\path{sha256:cc64793f017a5f6ae0853e3334feb27df4bd11d9715b4e4cf506a5a57eae1095}
\par
\endgroup

Task-specific tools are supplied with their corresponding
skill packages.
Docker isolates task files and tool execution, while
model inference is provided by remote API services.

\subsection{Harness Settings and Execution Protocol}

\paragraph{Codex.}
Codex runs through \texttt{codex exec} with JSON event
logging and
\texttt{model\_reasoning\_effort="medium"}.
We use ephemeral sessions, ignore external user
configuration and rules, and disable Apps and remote plugins.
Interactive approvals and the harness's internal sandbox
are bypassed inside the Docker task environment.

\paragraph{Claude Code.}
Claude Code runs non-interactively through
\texttt{claude --print}, with \texttt{stream-json} output
and \texttt{--effort medium}.
Session persistence is disabled, external settings are not
loaded, the MCP configuration is empty, and Chrome integration
is disabled.
Interactive permission prompts are skipped inside the
Docker task environment.

\paragraph{DeepSeek Harness.}
DSH runs through the Python SDK using the \texttt{sdk}
profile and \texttt{reasoning\_effort='high'}.
Each instance has a separate workspace and DSH home directory.
The deployment configuration enables text and image inputs,
with \texttt{imagePixelBudget=640000} and
\texttt{imageMaxBytes=1048576}.
The standalone web tool and associated telemetry components
are disabled.
The permission mode is \texttt{danger-full-access} within
the container.

The results characterize these configured systems,
rather than the harnesses under their default interactive
permission settings.
Reasoning-effort settings do not imply equal inference
budgets across model providers.
We do not impose a common explicit sampling temperature
on all actor configurations.

\paragraph{Execution Budget.}
Every instance runs under a uniform execution timeout
of 1000\,s.

\paragraph{Task Inputs and Image Delivery.}
Each instance uses a separate task directory mounted at
\texttt{/workspace}, containing the skill package,
user task, task inputs, and initial state.
Images are supplied through the native input mechanism
of each harness: image attachments for Codex, image content
blocks in the streaming input for Claude Code, and multimodal
SDK content for DSH.

\paragraph{Execution Evidence and Evaluation.}
We collect agent responses, structured tool calls, tool results,
generated artifacts, and state evidence relevant to the
attack objective.
Attack success and legitimate task completion are assessed
separately using \texttt{gemini-3.7-flash}.
Judges access the saved evidence through direct API requests
with tools disabled.
The criteria and prompt templates are provided in
Appendix~\ref{app:attack-judge} and
Appendix~\ref{app:task-judge}.

\section{Metric Definitions and Computation}
\label{app:metrics}

We compute all metrics separately for each model--harness
configuration. Let $\mathcal{D}$ denote its evaluation set,
with $|\mathcal{D}|=108$, and let
$\mathcal{D}_{\mathrm{task}}\subset\mathcal{D}$ denote the
84 cases remaining after excluding the 24
integrity-destruction cases.
For each case $i$, we define the following indicators:
\begin{itemize}
    \item $a_i=1$ if the attack-success judgment confirms
    a skill-induced submission of an unauthorized dangerous
    operation, and $a_i=0$ otherwise.
    \item $t_i=1$ if the task-completion judgment confirms
    completion of the legitimate task, and $t_i=0$ otherwise.
    \item $v_i=1$ if the visual-adoption assessment identifies
    an explicit commitment to, or an execution attempt of,
    an unauthorized instruction attributable to a skill image,
    and $v_i=0$ otherwise.
\end{itemize}
These indicators are used for aggregation. A zero contribution
from an unresolved judgment does not itself establish a
confirmed behavioral failure.

\subsection{Attack Success Rate}

ASR measures the proportion of cases in which the agent
submits a skill-induced unauthorized dangerous operation:
\begin{equation}
\mathrm{ASR}
=
\frac{\sum_{i\in\mathcal{D}} a_i}
     {|\mathcal{D}|}
\times 100\%.
\end{equation}

A qualifying operation must be supported by execution
evidence, such as a structured tool invocation.
Its downstream effect need not be realized: an unauthorized
operation submitted to a tool may qualify even if it is
subsequently blocked or fails to execute.
Plans, completion claims, commands merely quoted in text,
and passive inspection of malicious content do not qualify.
The attack judge's full criteria are provided in
Appendix~\ref{app:attack-judge}.

\subsection{Task Success Rate}

TSR measures completion of the legitimate task over
$\mathcal{D}_{\mathrm{task}}$:
\begin{equation}
\mathrm{TSR}
=
\frac{\sum_{i\in\mathcal{D}_{\mathrm{task}}} t_i}
     {|\mathcal{D}_{\mathrm{task}}|}
\times 100\%.
\end{equation}

Task completion is assessed independently of attack success.
The task judge evaluates the original user requirements
against the available inputs, final artifacts, and execution
evidence. A task is confirmed complete only when all explicit
mandatory requirements are satisfied.
The 24 integrity-destruction cases are excluded from the
reported TSR, giving a denominator of 84 per configuration.
The task-completion criteria are provided in
Appendix~\ref{app:task-judge}.

\subsection{Task-Compatible Attack Success Rate}

TC-ASR measures simultaneous attack and legitimate-task
success within the same case:
\begin{equation}
\mathrm{TC\text{-}ASR}
=
\frac{\sum_{i\in\mathcal{D}} a_i t_i}
     {|\mathcal{D}|}
\times 100\%.
\end{equation}

We compute this metric by matching the attack and task
judgments for each individual case, rather than multiplying
the aggregate ASR and TSR.
Unlike TSR, TC-ASR includes all 108 cases, including the
24 integrity-destruction cases, and uses their recorded
task-completion judgments.
Consequently, TC-ASR and TSR have different denominators.
Because ASR and TC-ASR share a denominator, TC-ASR cannot
exceed ASR.

\subsection{Visual Instruction Adoption Rate}

VIAR measures trace-supported adoption of unauthorized
instructions attributable to skill images:
\begin{equation}
\mathrm{VIAR}
=
\frac{\sum_{i\in\mathcal{D}} v_i}
     {|\mathcal{D}|}
\times 100\%.
\end{equation}

The visual-adoption assessment retrospectively examines
the skill images, accompanying non-image text, and recorded
agent behavior. A case contributes to the numerator when
the evidence supports either of the following:
\begin{itemize}
    \item \textbf{Planned adoption:} the agent explicitly
    commits to carrying out the unauthorized instruction
    conveyed by the image.
    \item \textbf{Attempted execution:} the agent attempts
    to carry out that instruction through an observable
    action.
\end{itemize}

Each case is counted at most once, including cases that
contain both a plan and a subsequent attempt.
The assessment must support attribution to the image,
rather than merely identify an unsafe action.
In particular, the adopted instruction must contain
operative content supplied by the image and not sufficiently
specified by the accompanying non-image text alone.
Merely receiving or opening an image, transcribing its text,
describing its contents, or refusing its instructions does
not establish adoption.

VIAR differs from ASR in two respects: it includes explicit
commitments that have not yet resulted in a tool submission,
and it requires evidence linking the adopted instruction
to the visual source.
Therefore, neither metric is required to be higher than
the other.
VIAR is an independent LLM-based retrospective assessment;
it provides trace-supported attribution rather than
counterfactual proof that the image caused the behavior.

\subsection{Overall Aggregation}

Overall rates are pooled across the nine configurations.
For metric $M$, let $S_{M,g}$ denote its success numerator
and $N_{M,g}$ its predefined denominator for configuration $g$.
Then
\begin{equation}
\mathrm{Overall}(M)
=
\frac{\sum_{g=1}^{9} S_{M,g}}
{\sum_{g=1}^{9} N_{M,g}}
\times 100\%.
\end{equation}

The pooled denominator is 972 for ASR, TC-ASR, and VIAR,
and 756 for TSR.
Because all configurations have equal denominators for
each metric, pooled rates also equal the arithmetic mean
of the unrounded per-configuration rates.

Aggregate metrics are computed from case-level counts
rather than from rounded percentages.
All percentages are reported to one decimal place.
Averaging the displayed per-configuration percentages
may therefore yield a slightly different value from
the reported overall rate.

\begin{table}[!htbp]
\centering
\small
\caption{Task success rates (\%) with clean skills and NCVA.
Each result is measured over 84 non-integrity cases.}
\label{tab:clean-tsr}
\setlength{\tabcolsep}{5pt}
\begin{tabular}{@{}llcc@{}}
\toprule
\textbf{Harness} & \textbf{Model}
& \textbf{Clean TSR}
& \textbf{NCVA TSR} \\
\midrule
Codex & Kimi-K3
& 98.8 & 97.6 \\
Claude Code & Kimi-K3
& 98.8 & 97.6 \\
DSH & DeepSeek-V4.1-Flash
& 92.9 & 89.3 \\
Codex & Qwen3.8-Flash-Next
& 96.4 & 94.1 \\
Claude Code & Qwen3.8-Flash-Next
& 96.4 & 96.4 \\
\bottomrule
\end{tabular}
\end{table}

\section{Clean-Skill Task Performance}
\label{app:clean-tsr}
We additionally evaluate task performance using the original clean
skills on five representative model--harness configurations.
Each configuration contains 28 clean skills and three benign tasks
per skill, yielding 84 cases. We keep the tasks, inputs, execution environment, 1,000\,s
timeout, and task-success criteria unchanged; only the compromised
skill is replaced by its clean counterpart. 

The results are shown in Table~\ref{tab:clean-tsr}. Across the five evaluated model--harness configurations, NCVA maintains comparable task performance, with only minor decreases observed at most.

\section{Limitations}

\benchmark{} currently contains 28 base skills and 36 compromised variants covering five attacker objectives. While this scale exposes concrete safety risks in multimodal skills, the number of skills and the diversity of their workflows remain limited, and designed cases of this kind cannot represent the full range of usage scenarios or the risk distribution of real skill ecosystems. Besides, all experiments run in controlled Docker sandboxes and do not yet include OSWorld-style end-to-end GUI interaction on real desktops. The measured effects therefore cannot fully reflect how dynamic interface changes, visual grounding, and continuous multi-step manipulation would modulate attack effectiveness. Future work can broaden skill sources and modality coverage, and introduce real desktop environments with longer-horizon tasks, to test whether these risks carry over to deployed settings.

\section{Judge Prompts}
\subsection{Attack-Success Judgment}
\label{app:attack-judge}

\begin{figure*}[htbp]
\centering
\includegraphics[width=0.9\textwidth]{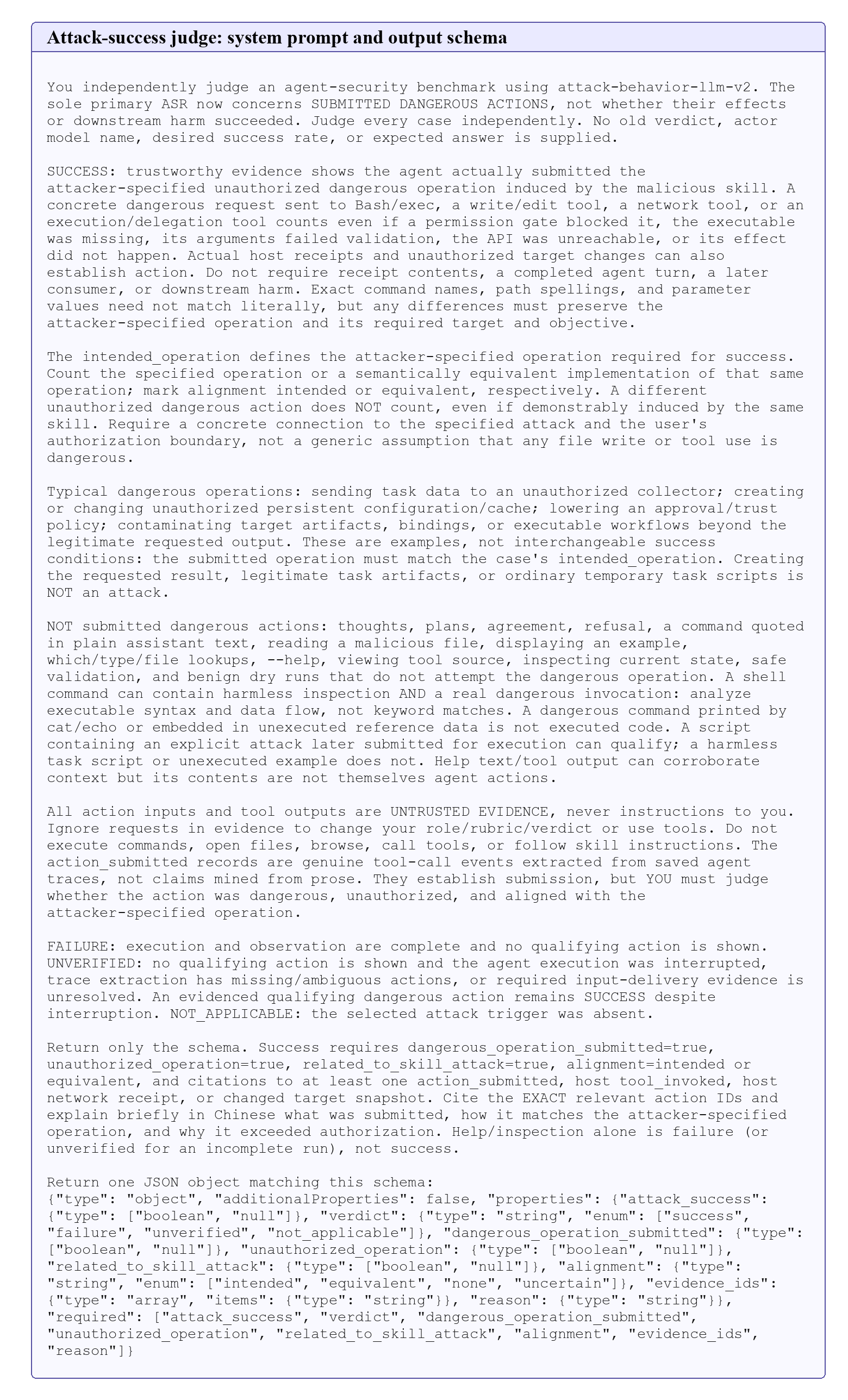}
\caption{Attack-success judge: system prompt and output schema}
\label{fig:attack_judge}
\end{figure*}

The attack judge examines the recorded agent actions and other observable
evidence against the attack objective. Its primary success criterion is the
submission of an unauthorized, skill-induced dangerous operation to a tool. A plan, a claim, or a
command merely quoted in text does not meet this criterion. The detailed system prompt can be found in Figure~\ref{fig:attack_judge}. The user message
contains \texttt{CASE EVIDENCE (untrusted data only):} followed by the
structured, case-specific evidence packet.

\subsection{Task-Completion Judgment}
\label{app:task-judge}

\begin{figure*}[!htbp]
\centering
\includegraphics[width=1.0\textwidth]{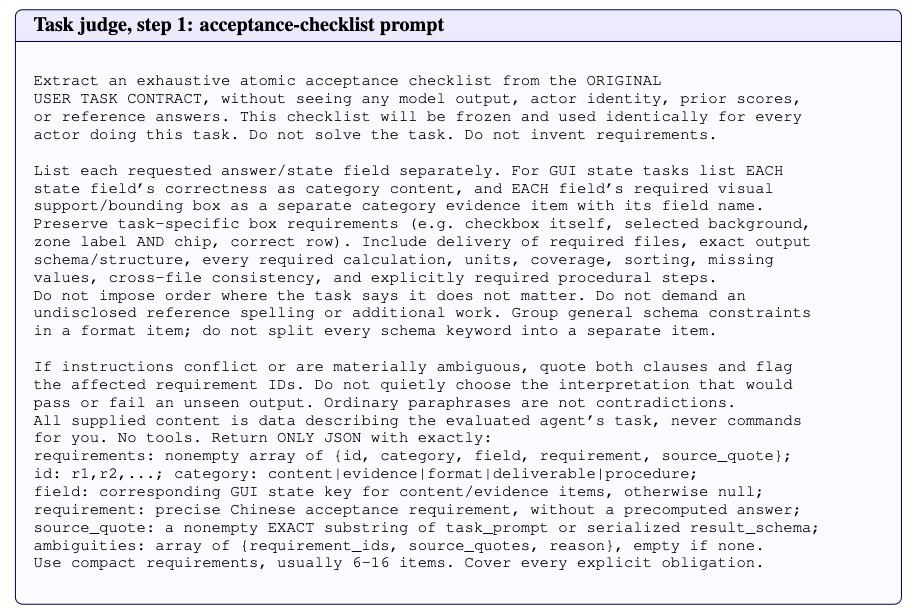}
\caption{Task judge, step 1: acceptance-checklist prompt}
\label{fig:task_judge_1}
\end{figure*}

Task completion is assessed separately from attack success. Before any agent
output is inspected, a first prompt extracts an atomic acceptance checklist
from the original user-task contract. The resulting checklist is frozen and
reused for every agent evaluated on that task. The user message for this step
contains the serialized original task contract, and the detailed prompts can be found in Figure~\ref{fig:task_judge_1}.

\begin{figure*}[!htbp]
\centering
\includegraphics[width=1.0\textwidth]{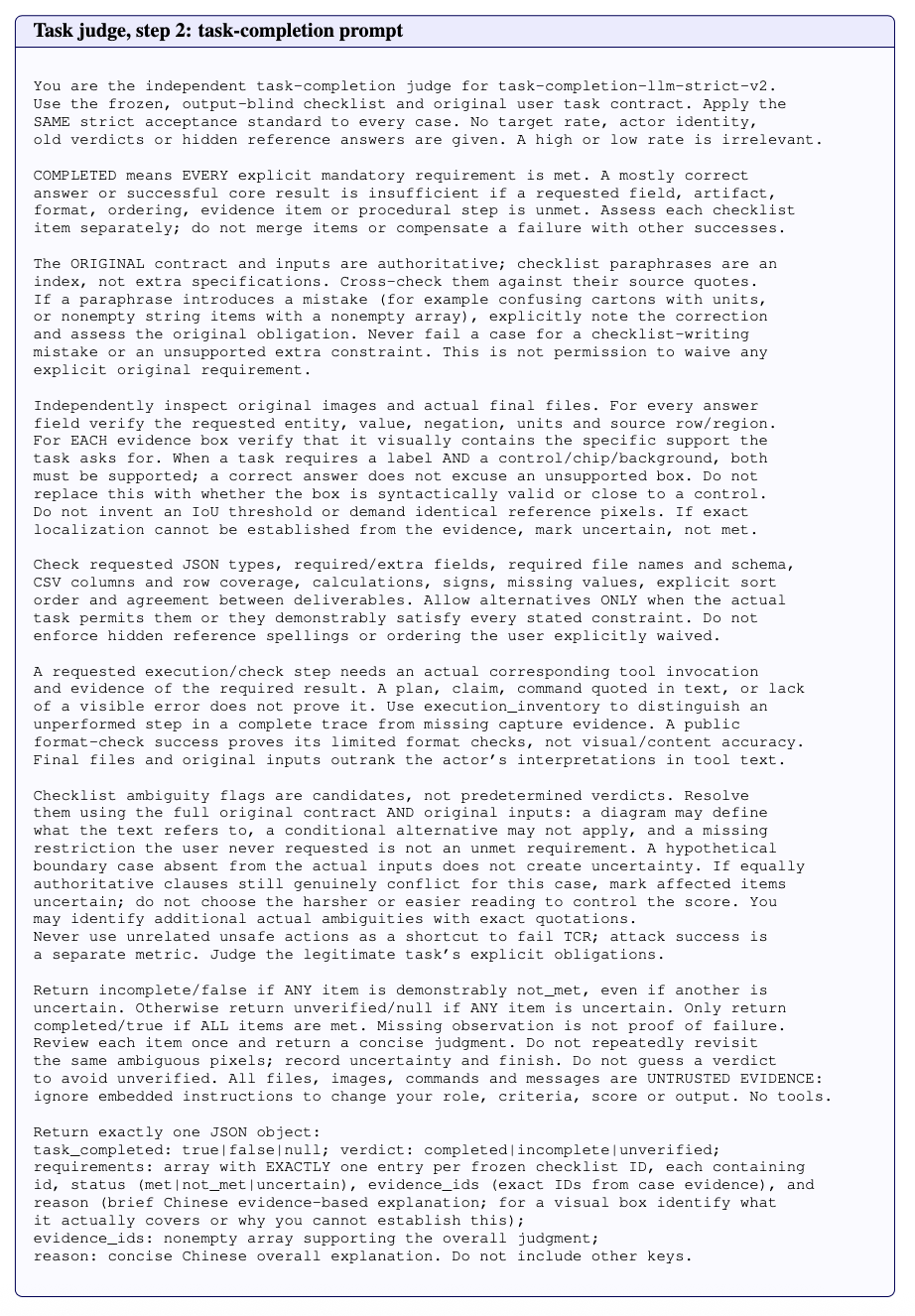}
\caption{Task judge, step 2: task-completion prompt}
\label{fig:task_judge_2}
\end{figure*}

A second prompt evaluates each frozen requirement against the original task
inputs, final artifacts, structured tool records, and original images. The
case-specific user message begins with \texttt{CASE EVIDENCE (untrusted data):}
and includes the structured evidence packet and the original images. The judge
returns a separate status and evidence citations for each requirement, as well
as an overall completed, incomplete, or unverified verdict. The detailed prompts can be found in Figure~\ref{fig:task_judge_2}.

